\documentclass{emulateapj}

\usepackage{apjfonts}

\newcommand{\cool}{\mathrm{cool}}
\newcommand{\ext}{\mathrm{ext}}
\newcommand{\diff}{\mathrm{diff}}

\begin{document}

\title{Did Fomalhaut, HR~8799, and HL~Tauri Form Planets via the Gravitational
Instability?\\
Placing Limits on the Required Disk Masses}

\author{D. Nero \& J. E. Bjorkman}

\affil{Ritter Observatory, MS 113, Department of Physics and Astronomy,
University of Toledo, Toledo, OH 43606-3390}

\email{dnero@physics.utoledo.edu}
\begin{abstract}
Disk fragmentation resulting from the gravitational instability has
been proposed as an efficient mechanism for forming giant planets.
We use the planet Fomalhaut~b, the triple-planetary system HR~8799,
and the potential protoplanet associated with HL~Tau to test the
viability of this mechanism. We choose the above systems since they
harbor planets with masses and orbital characteristics favored by
the fragmentation mechanism. We do not claim that these planets \emph{must}
have formed as the result of fragmentation, rather the reverse: if
planets can form from disk fragmentation, then these systems are consistent
with what we should expect to see. We use the orbital characteristics
of these recently discovered planets, along with a new technique to
more accurately determine the disk cooling times, to place both lower
and upper limits on the disk surface density---and thus mass---required
to form these objects by disk fragmentation. Our cooling times are
over an order of magnitude shorter than those of \citet{rafikov05},which
makes disk fragmentation more feasible for these objects. We find
that the required mass interior to the planet's orbital radius is
$\sim0.1\ M_{\sun}$ for Fomalhaut~b, the protoplanet orbiting HL~Tau,
and the outermost planet of HR~8799. The two inner planets of HR~8799
probably could not have formed \emph{in situ} by disk fragmentation. 
\end{abstract}


\section{Introduction}

As more extrasolar planets are discovered, we are increasingly pressed
to describe how planets can form in such a variety of environments.
Until just recently, observational selection biases have resulted
in the fact that all observed extrasolar planets have been found to
orbit within a few AU of their star \citep{butler06}. Since it seems
unlikely that these planets could have formed \emph{in situ} \citep{mayor95},
planet migration is usually invoked \citep{alibert05}. Unfortunately,
this means that little is known about where---and hence how---these
planets originally formed.

In contrast, the technique of direct-imaging has presented us with
a new set of extrasolar planets that lie far from their star \citep{kalas08,marois08},
along with a potential protoplanet \citep{greaves08}. Like previous
techniques, direct imaging preferentially detects giant planets of
several Jupiter masses. Furthermore, planet migration need not be
invoked to explain how these planets could form at their observed
locations.

One possible mechanism for giant planet formation is core accretion
followed by rapid gas accretion \citep{pollack96,inaba03}. However,
this mechanism has difficulty forming giant planets at large radii.
The primary reason for this is that the initial core accretion time
scales as $r^{3}$, where $r$ is the orbital radius of the planet
\citep{ikoma00,kenyon08}. Thus, while it may take $\sim1$~Myr to
form a gas giant at 5~AU via core accretion, it would take $\sim1$~Gyr
for the same process at 50~AU---far longer than the observed lifetimes
of protoplanetary disks \citep{haisch01}.

Another mechanism for giant planet formation is disk fragmentation
as a consequence of the gravitational instability \Citep[see also the recent review by \citealt{durisen07} and \citealt{stamatellos09} for recent developments]{kuiper51,cameron78,boss97}.
Provided that the disk surface density is sufficiently large, this
mechanism can form giant planetary embryos on time scales of a few
orbital periods. However, if the surface density is too large, the
disk is unable to cool sufficiently fast for fragmentation to take
place at all \citep{rafikov05}. The combination of these requirements
implies gravitational instability can only form massive planets at
large radii.

In this letter, we consider the planet \object{Fomalhaut~b} \citep{kalas08},
the triple-planet system \object{HR~8799} \citep{marois08}, and
the potential protoplanet orbiting \object{HL~Tau}%
\footnote{For our purposes, the distinction between planet and protoplanet is
irrelevant, and we will use {}``planet'' in both contexts from here
on.%
} \citep{greaves08}. Each of these systems possesses at least one
planet with orbital characteristics favored by the disk fragmentation
mechanism. By determining the range of surface densities required
to form a giant planet with the same semi-major axis as these observed
planets, we can infer the range of disk masses needed for the fragmentation
mechanism to have operated in these systems.

\section{Disk Fragmentation}

The stability of a thin, massive disk is controlled by the \citet{toomre64}
$Q$ parameter \begin{equation}
Q\equiv\frac{c_{s}\Omega}{\pi G\Sigma}\ ,\label{eq:Q}\end{equation}
 where $c_{s}$ is the isothermal sound speed, $\Omega$ is the orbital
angular frequency (assuming a Keplerian disk), and $\Sigma$ is the
surface density. The disk becomes gravitationally unstable for $Q\lesssim1$.
However, even if a disk is gravitationally unstable, it can only fragment
if it possesses a sufficiently short cooling time \citep{gammie01,rice03}.
Specifically, fragmentation will only occur if \begin{equation}
t_{\cool}<\frac{\xi}{\Omega}\ ,\label{eq:tcool-Omega}\end{equation}
 where $t_{\cool}$ is the local cooling time for a small, point-source
perturbation, and $\xi$ is a factor of order unity that can depend
on the history of the disk \citep{clarke07}. We adopt $Q<1$ and
$\xi=1$ for our fragmentation criteria.

\subsection{Local Cooling Time}

Typically, the effects of cooling have been studied using time-dependent
hydrodynamic simulations. Inevitably, these numerical approaches have
to employ significant simplification of the radiation field for the
sake of computation time (e.g., optically thin cooling or flux limited
diffusion). Many of the simulations show that fragmentation does occur
given sufficiently high surface densities \citep{durisen07}.

In contrast, \citet{rafikov05} used an analytic, order-of-magnitude
calculation to show that cooling times derived from the equations
of radiative transfer were much longer, and that fragmentation thus
did not work, except at radii $\gtrsim100$~AU. Here, we adopt an
approach inspired by Rafikov, but with a more complete calculation
of the radiative transfer. In brief, we find cooling times that are,
in most cases, over an order of magnitude shorter than those given
by Rafikov (see Nero \& Bjorkman, in prep.\ for a more complete discussion).
As a consequence, we find that fragmentation over a larger range of
the outer disk is possible, depending on the details of the system.

We emphasize that the cooling time we calculate here is for a \emph{perturbation},
and is not the same as the total disk cooling time employed by \citet{gammie01}.
While the later may be more convenient for numerical hydrodynamic
simulations, the former is necessary to properly account for background
heating by external illumination (i.e., the host star). The perturbation
cooling time determines the onset and initial growth of the instability
(in the linear regime), while the total cooling time controls the
ultimate (typically non-linear) completion of the instability. Note,
however, that when self-heating is small, the perturbation and total
cooling times will be the same within a factor of order unity.

The perturbation cooling time $t_{\cool}=\Delta\mathcal{E}/(8\pi\Delta H_{0})$,
where $\Delta\mathcal{E}$ is energy per unit area added by the perturbation,
and $\Delta H_{0}$ is the frequency-integrated Eddington flux at
the disk surface. We consider an annulus within the disk, which we
approximate as a plane-parallel atmosphere with finite thickness.
For simplicity, we assume that the perturbation is located at the
disk mid-plane and that the disk cools equally from its top and bottom
surfaces. Under these assumptions, the perturbation cooling time is
\begin{equation}
t_{\cool}=\frac{1}{16}\frac{c_{{\rm m}}^{2}}{\gamma_{A}-1}\frac{1}{\sigma T_{{\rm m}}^{4}}\frac{1}{\chi^{\diff}}\int_{0}^{\tau_{0}}\left(\frac{B}{B_{{\rm m}}}\right)^{-3/4}\frac{\Delta B}{\Delta H_{0}}\,{\rm d}\tau\ ,\label{eq:tcool}\end{equation}
where $\gamma_{A}$ is the adiabatic constant for the gas, $\chi^{\diff}$
is the mean opacity (absorption plus scattering), $\tau$ is the optical
depth coordinate, $B$ and $\Delta B$ are the depth-dependent Planck
function and its perturbation, and $B_{{\rm m}}$, $T_{{\rm m}}$,
and $c_{m}$ are the Planck function, the temperature, and the isothermal
sound speed at the disk mid-plane, respectively. The limits of integration,
$\tau_{0}$ and $-\tau_{0}$, are the optical depth coordinates at
the {}``top'' and {}``bottom'' surface, respectively. Note that
we break convention here by placing $\tau=0$ at the disk mid-plane,
rather than at the top surface. Locations in the disk below the disk
mid-plane have negative optical depth coordinates, while those above
have positive. Also, note that we have assumed that $\chi^{\diff}$
is approximately constant over the vertical extent of the disk, since
the disk is nearly isothermal in the vertical direction. While this
is not true for the surface layers, the error is minimal since most
of the disk mass---and thus internal energy---is located in the disk
interior. Similarly, the vertically isothermal assumption does not
apply when accretion is the dominant source of heating (since then
there would be a significant vertical temperature gradient); however,
at large radii accretion luminosity is usually not the dominant heating
mechanism for the disk.

We assume that the relevant physics (i.e., reprocessing/absorption
of external radiation along with viscous energy generation) can be
preserved by splitting the intensity into two frequency components:
\emph{diffuse}, which corresponds to photons that have been reprocessed
and emitted by the disk, and \emph{external}, which corresponds to
unabsorbed photons emitted from the central star and potentially scattered
in the disk. Assuming gray opacity (i.e., the appropriate mean for
each spectrum), the frequency integrated moments of the transfer equations
are\begin{eqnarray}
\frac{dH^{\ext}}{d\tau} & = & -\frac{\kappa^{\ext}}{\chi^{\diff}}J^{\ext}\label{eq:1st-ext}\\
\frac{dJ^{\ext}}{d\tau} & = & -\frac{\chi^{\ext}}{\chi^{\diff}}\frac{H^{\ext}}{f^{\ext}}\label{eq:2nd-ext}\\
\frac{dH^{\diff}}{d\tau} & = & \frac{\kappa^{\diff}}{\chi^{\diff}}\left(B-J^{\diff}\right)\label{eq:1st-diff}\\
\frac{dJ^{\diff}}{d\tau} & = & -\frac{H^{\diff}}{f^{\diff}}\ ,\label{eq:2nd-diff}\end{eqnarray}
where $J$ is the mean intensity, $H$ is the Eddington flux, and
$\chi$ and $\kappa$ are the total and absorptive opacity, respectively.
We have written all optical depth coordinates in terms of $\tau\equiv\tau^{\diff}=\tau^{\ext}\chi^{\diff}/\chi^{\ext}$.
The Eddington factors, $f^{\ext}=K^{\ext}/J^{\ext}$ and $f^{\diff}=K^{\diff}/J^{\diff}$,
are assumed to be constant with depth. Note that we have included
no thermal emission for the \emph{external} frequency because the
disk is typically much cooler than the star.

To find the \emph{external} radiation, we combine eqs.~(\ref{eq:1st-ext})
and (\ref{eq:2nd-ext}), which have the solution\begin{equation}
H^{\ext}=H_{0}^{\ext}\frac{\sinh\beta\tau}{\sinh\beta\tau_{0}}\ ,\label{eq:Hext}\end{equation}
where $\beta\equiv\kappa^{\ext}\chi^{\ext}/[(\chi^{\diff})^{2}f^{\ext}]$
and $H_{0}^{\ext}$ is the net \emph{external} surface flux.

There are three sources of energy for the \emph{diffuse} radiation:
1) absorption of \emph{external} radiation $-dH^{\ext}/d\tau$, 2)
accretion luminosity $L_{\mathrm{acc}}$ with surface flux $H_{0}=(dL_{\mathrm{acc}}/dA)/8\pi$,
and 3) the point-source perturbation at the mid-plane $\Delta H_{0}\delta(\tau)$.
Thus, the flux transported by the disk is $H^{\diff}=H_{0}(\tau/\tau_{0})+\Delta H_{0}\mathrm{sgn}\,\tau-H^{\ext}$.

From eq.~(\ref{eq:2nd-diff}), we can now obtain the \emph{diffuse}
mean intensity \begin{eqnarray}
J^{\diff} & = & H_{0}\left(\frac{\tau_{0}^{2}-\tau^{2}}{2f^{\diff}\tau_{0}}+\frac{1}{g^{\diff}}\right)+\Delta H_{0}\left(\frac{\tau_{0}-|\tau|}{f^{\diff}}+\frac{1}{g^{\diff}}\right)\nonumber\\
 & - & H_{0}^{\ext}\left(\frac{\cosh\beta\tau_{0}-\cosh\beta\tau}{\beta f^{\diff}\sinh\beta\tau_{0}}+\frac{1}{g^{\diff}}\right)\ ,\label{eq:Jdiff}\end{eqnarray}
where $g^{\diff}=H^{\diff}(\tau_{0})/J^{\diff}(\tau_{0})$ is the
second Eddington factor.

Finally, the temperature may now be found from radiative equilibrium,
eq.~(\ref{eq:1st-diff}), which gives \begin{equation}
B=J^{\diff}+\frac{\chi^{\diff}}{\kappa^{\diff}}\frac{dH^{\diff}}{d\tau}\ ,\label{eq:planck-fn}\end{equation}
as well as\begin{equation}
\Delta B=\Delta H_{0}\left(\frac{\tau_{0}-|\tau|}{f^{\diff}}+\frac{1}{g^{\diff}}+2\frac{\chi^{\diff}}{\kappa^{\diff}}\delta(\tau)\right)\ .\end{equation}

Given the temperature and its perturbation, we calculate the cooling
time from eq.~(\ref{eq:tcool}). In terms of the surface density,
$\Sigma=2\tau_{0}/\chi^{\diff}$, the cooling time \begin{equation}
t_{\cool}\propto\Sigma^{a}\ ,\end{equation}
where the exponent $a$ depends on both the disk optical depth and
the ratio of heating from the star to self-heating from accretion
(see Nero \& Bjorkman, in prep.), but is generally in the range 0--2.
In the negligible self-heating regime ($H_{0}\rightarrow0$, $B/B_{m}\rightarrow1$),
eq.~(\ref{eq:tcool}) reduces to \begin{equation}
t_{\cool}=\frac{1}{16}\frac{c_{{\rm m}}^{2}}{\gamma_{A}-1}\frac{1}{\sigma T_{{\rm m}}^{4}}\left(\frac{\chi^{\diff}\Sigma^{2}}{8f^{\diff}}+\frac{\Sigma}{2g^{\diff}}+\frac{1}{\kappa^{\diff}}\right)\ .\label{eq:tcool-thick+thin}\end{equation}
In the optically thick limit ($\tau_{0}\gg1,$ $f^{\diff}=1/3$),
eq.~(\ref{eq:tcool-thick+thin}) further simplifies to \begin{equation}
t_{\cool}\approx\frac{3}{128}\frac{c_{{\rm m}}^{2}}{\gamma_{A}-1}\frac{1}{\sigma T_{{\rm m}}^{4}}\chi^{\diff}\Sigma^{2}\ .\label{eq:tcool-thick}\end{equation}
Note that this expression is directly comparable to the optically
thick limit of eq.~(4) in \citet{rafikov05}, with the exception
that our numerical factor (3/128) is over an order of magnitude smaller
than his (1/4), thus leading to significantly reduced cooling times.
This difference from Rafikov's result is much larger than the difference
between total and perturbation cooling times. It instead arises from
his assumption that several factors were of order unity, which is
true individually; however, the cumulative effect of his assumptions
results in his making a large overestimate. 

Our analytic expression for the cooling time, eq.~(\ref{eq:tcool-thick}),
assumes no accretion luminosity and large optical depth. While it
is probably safe to assume that self-heating is small at large disk
radii, it is not as certain that the optically thick limit will hold.
For this reason, we use the full solution of eq.~(\ref{eq:tcool})
when calculating $t_{\cool}$, rather than assuming that this limit
holds. Nonetheless, eq.~(\ref{eq:tcool-thick}) makes for a good
estimate in most cases where $\Sigma$ is large enough to satisfy
$Q<1$ in the outer disk.

\subsection{Fragment Masses}

When determining if disk fragmentation is a viable mechanism for forming
giant planets, another point to consider is the issue of producing
the proper planetary mass of a few $M_{{\rm Jupiter}}$. While a full
treatment of this problem is beyond the scope of this letter, we provide
a toy model to argue that this is likely to be the case.

If $Q\lesssim1$ at some radius $r$, the disk becomes gravitationally
unstable. Supposing $m$ spiral arms form, each arm has local surface
density\begin{equation}
\Sigma_{\mathrm{arm}}=\frac{w}{m}\frac{r}{R}\Sigma\ ,\end{equation}
where $\Sigma$ is the original surface density of the previously
axisymmetric disk, $R$ is the current width of the spiral arms, and
$w$ is a constant that depends on the winding angle. For simplicity,
we assume that most of the disk mass is confined to the spiral arms,
while the space between the arms is effectively empty. In addition,
we assume $w\sim1$, corresponding to moderate winding.

If $t_{\cool}\gtrsim1/\Omega$, then the spiral arms are pressure
supported and are stable against fragmentation. However, if $t_{\cool}\lesssim1/\Omega$,
then the arms are instead supported by centrifugal forces. As $R$
continues to decrease, they will fragment radially once the centrifugal
support is lost. Balancing self-gravity against the centrifugal support,
$\Omega^{2}R=\pi G\Sigma_{\mathrm{arm}}$, fragmentation occurs when
\begin{equation}
R<R_{f}\equiv r^{2}\sqrt{\frac{\pi}{m}\frac{\Sigma}{M_{\star}}}\ ,\end{equation}
where $M_{\star}$ is the stellar mass. The fragment mass is $\pi R_{f}^{2}\Sigma_{\mathrm{arm}}$,
so assuming $m\sim\pi$, we find a characteristic fragment mass \begin{equation}
M_{f}\sim1M_{{\rm Jupiter}}\left(\frac{\Sigma}{10{\rm gcm}^{-2}}\right)^{3/2}\left(\frac{M_{\star}}{M_{\sun}}\right)^{-1/2}\left(\frac{r}{100{\rm AU}}\right)^{3}\ ,\label{eq:mass}\end{equation}
which is consistent with our requirement to produce Jupiter-mass planets.

\section{Disk Mass Limits}

The condition that both the Toomre $Q$ and the cooling time be sufficiently
small (eqs.\ {[}\ref{eq:Q}{]} and {[}\ref{eq:tcool-Omega}{]}) can
be used to place limits on what disk surface densities lead to fragmentation.
To be gravitationally unstable, the Toomre $Q$ condition requires
that the surface density be larger than a minimum, $\Sigma_{\min}$,
while the cooling condition imposes a maximum surface density, $\Sigma_{\max}$.
Therefore, local disk fragmentation is only possible if $\Sigma_{\min}<\Sigma<\Sigma_{\max}$.
It of course follows that for fragmentation to be possible at all,
$\Sigma_{\min}$ must be less than $\Sigma_{\max}$. This limits the
range of radii were fragmentation is even a possibility.

\begin{figure}
\includegraphics[clip]{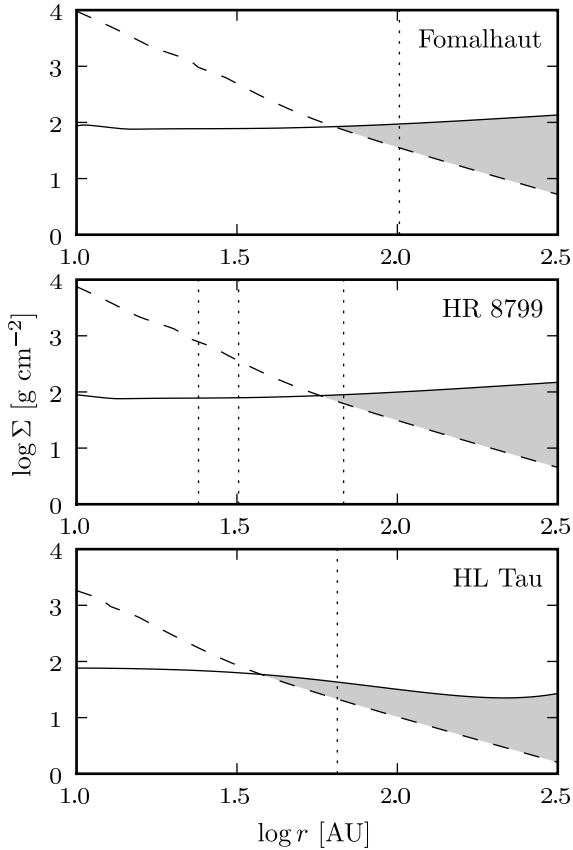} 

\caption{\label{fig:2much2little}Surface density limits for disk fragmentation.
The solid line denotes $\Sigma_{\max}$, which is the maximum surface
density for the cooling time constraint. The dashed line denotes the
minimum surface density for fragmentation, $\Sigma_{\min}$, which
is the locus Toomre $Q=1$. Disk fragmentation is only allowed in
the region $\Sigma_{\max}>\Sigma_{\min}$, which is shaded gray. The
locations of known planets are plotted as vertical dotted lines.}

\end{figure}

\begin{deluxetable}{lcccr}
\tablecaption{\label{tab:parameters}Parameters Used for Each System}
\tablehead{
\colhead{Object} & \colhead{$M_{\star}$} & \colhead{$R_{\star}$} & \colhead{$T_{\star}$} & \colhead{$r_{\mathrm{planet(s)}}$}\\
\colhead{} & \colhead{$(M_{\sun})$} & \colhead{$(R_{\sun})$} & \colhead{(K)} & \colhead{(AU)}
}
\startdata
Fomalhaut b & 2.0 & 1.8 & 8500 & $>101.5$\tablenotemark{a}\\
HR 8799 b, c, d & 1.5 & 1.8 & 8200 & 68, 38, 24\tablenotemark{b}\\
HL Tau b & 0.3 & 0.6 & 3700 & 65\tablenotemark{c}\\
\enddata
\tablenotetext{a}{\citet{chiang08}}
\tablenotetext{b}{\citet{marois08}}
\tablenotetext{c}{\citet{greaves08}}
\end{deluxetable}

\begin{deluxetable*}{lcccccc}
\tablecaption{\label{tab:Mass-Limits}Range of Disk Masses that Fragment}
\tablehead{
\colhead{Object} &
\colhead{$\Sigma$} &
\colhead{$M_d\ (p=0.0)$} &
\colhead{$M_d\ (p=0.5)$} &
\colhead{$M_d\ (p=1.0)$} &
\colhead{$M_d\ (p=1.5)$} &
\colhead{$M_f$}\\
\colhead{} &
\colhead{(g cm$^{-2}$)} &
\colhead{$(M_{\sun})$} &
\colhead{$(M_{\sun})$} &
\colhead{$(M_{\sun})$} &
\colhead{$(M_{\sun})$} &
\colhead{$(M_{\mathrm{Jupiter}})$}
}
\startdata
Fomalhaut b & 35--94 & 0.13--0.34 & 0.17--0.45 & 0.26--0.68 & 0.51--1.36 & 5--21\\
HR 8799 b   & 62--89 & 0.10--0.15 & 0.14--0.19 & 0.20--0.29 & 0.41--0.58 & 4--7\\
HL Tau b    & 21--43 & 0.03--0.06 & 0.04--0.09 & 0.06--0.13 & 0.13--0.26 & 2--4\\
\enddata
\end{deluxetable*}

In Fig.~\ref{fig:2much2little}, we plot $\Sigma_{\min}$ and $\Sigma_{\max}$
for the systems Fomalhaut, HR~8799, and HL~Tau, using the parameters
listed in Table~\ref{tab:parameters}. In all cases we used an accretion
rate of $10^{-6}M_{\sun}{\rm yr}^{-1}$, a mean molecular weight for
the disk of $\mu_{g}=2.33$, and an adiabatic gas constant of $\gamma_{A}=1.43$.
The temperature is determined from eq.~(\ref{eq:planck-fn}), using
a flared disk model with a power law scale height $h\propto r^{5/4}$,
which determines the angle of incidence of the \emph{external} radiation.
We use the dust opacity of \citet{cotera01} and the Rosseland mean
gas opacity of \citet{helling00}. We note that the stellar parameters
in Table~\ref{tab:parameters} are not necessarily appropriate if
planet formation occurs during the Class 0/I phase when the star is
significantly more luminous. However, our results are relatively insensitive
to this effect, since $T_{m}$ is only weakly dependent on the stellar
luminosity.

Each system presented here has a planet that might have formed via
disk fragmentation, assuming that the local surface density had a
value between $\Sigma_{\min}$ and $\Sigma_{\max}$ for at least a
few orbital periods. By assuming a power law for the surface density,
$\Sigma\propto r^{-p}$, we can calculate a range of disk masses (interior
to $r_{\mathrm{planet}}$) that satisfies this condition.

A survey of 24 circumstellar disks by \citet{andrews07} found $p\approx$
0.0--1.0 with an average of $p\approx0.5$, while the hydrodynamical
simulations of \citet{vorobyov09b} found $p\approx$ 1.0--2.0 with
an average around $p\approx1.5$. Disk mass limits $M_{d}$ for $p=0.0$,
0.5, 1.0 and 1.5, along with the more fundamental surface density
limits are given in Table~\ref{tab:Mass-Limits}. We also provide
the characteristic fragment mass (approximate planet mass) $M_{f}$
from eq.~(\ref{eq:mass}) that we would expect from the disk fragmentation
mechanism. Also note that, in order to be conservative, we are using
the smallest radius found by \citet{chiang08} for Fomalhaut~b. Using
one of their better fits (e.g.\ 115 AU) will decrease our lower disk
mass limit by a few percent, increase our upper disk mass limit by
$\approx30\%$ (which would make fragmentation slightly easier), and
increase the characteristic fragment mass by $\approx50\%$.

\section{Discussion}

We have refined the calculations of \citet{rafikov05} and found cooling
times over an order of magnitude shorter. We have used these cooling
times, along with the observed stellar parameters of Fomalhaut, HR~8799,
and HL~Tau, to test the viability of the disk fragmentation mechanism.
We found that in each of these systems, at least one planet could
have formed \emph{in situ} as the result of fragmentation, assuming
the disk mass interior to those planets fell within a particular range
as indicated in Table~\ref{tab:Mass-Limits}.

While the ranges in Table~\ref{tab:Mass-Limits} only span a factor
of a few, this is not by itself a significant limitation. Even if
the local surface density is above the upper instability limit, fragmentation
may still occur since the surface density must eventually drop through
the unstable regime as the disk evolves and dissipates. The caveat
is that the surface density needs to evolve on a timescale longer
than an orbital period so that there can be sufficient time to fragment.

Our minimum disk masses for Fomalhaut~b, HR~8799\textbf{~}b, and
HL~Tau~b are about an order of magnitude larger than those inferred
from observations \citep{andrews07}. Note, however, that this is
a problem for all planet formation models in general. Even core accretion
models require an enhanced surface density (although to a somewhat
lesser extent) \citep{pollack96,inaba03}. One possible mechanism
for increasing the surface density is mass loading from an infalling
envelope \citep{vorobyov06}. Conversely, current estimates of disk
masses may be too low because they depend on: 1) the extrapolation
of surface densities in the outermost regions of the disk to the inner
disk, and 2) the rather uncertain dust opacity. For example, larger
dust grains would require larger disk masses to fit the observed SEDs
\citep{andrews07}.

As further evidence for underestimated disk masses, numerical hydrodynamical
simulations by \citet{vorobyov09a} found disk masses much higher
than those of \citet{andrews07}. In particular, stars like Fomalhaut
and HR~8799 can support disks as large as 0.5 $M_{\sun}$, while
HL~Tau could have a disk as massive as 0.1 $M_{\sun}$, all of which
are within our limits for disk fragmentation. We caution, however,
that our choice of opacity model can have a major effect on our results.
For example, decreasing the dust opacity raises the temperature and
decreases the cooling time in the outer disk, resulting in disk fragmentation
at smaller radii. On the other hand, increasing the opacity would
have the opposite effect.

Regardless of the above considerations, HR~8799~c and d are too
close to their parent star to have formed \emph{in situ} via fragmentation
under the conditions modeled here. Appealing to chronically overestimated
dust opacity can only get us so far. Dropping the opacity by an order
of magnitude brings HR~8799~c into the fragmentation zone, but still
leaves HR~8799~d out. Likewise, twiddling other parameters can also
move the fragmentation radius inward, but reaching the required 24~AU
with reasonable parameters does not seem possible. We therefore conclude
that HR~8799~c and d likely did not form \emph{in situ }as the result
of disk fragmentation (of course, they could have formed at larger
radii were the disk is more likely to fragment and migrated inward).

For those planets that could form by disk fragmentation, we find characteristic
fragment masses (approximate planet masses) of a few $M_{{\rm Jupiter}}$
for the lower end of unstable disk surface density. Our estimates
are mostly consistent with expectations, although \citet{chiang08}
found Fomalhaut~b to have an upper mass limit of 3~$M_{{\rm Jupiter}}$,
which is 60\% lower than our lowest estimate of 5~$M_{{\rm Jupiter}}$.
Nonetheless, considering the crudeness of eq.~(\ref{eq:mass}), we
are not convinced that this discrepancy rules out Fomalhaut\textbf{~}b
from having formed as a result of disk fragmentation.

In conclusion, given our current uncertainty of typical disk masses
and dust opacities, we cannot rule out planet formation from disk
fragmentation. Of the three currently known systems that might posses
planets formed in this manner, all are able to produce at least their
outermost giant planet via fragmentation, given a large enough disk
mass at some point in their history.

\acknowledgements{}

This research was supported by NSF Grant AST-0307686 to the University
of Toledo. We would like to thank Ken Rice and Lee Hartmann for their
valuable insights regarding this work. We also wish to thank the anonymous
referee for very helpful suggestions that significantly improved the
manuscript.


\begin{thebibliography}{}
\bibitem[Alibert et al.(2005)]{alibert05}Alibert, Y., Mordasini,
C., Benz, W., \& Winisdoerffer, C.\ 2005, \aap, 434, 343

\bibitem[Andrews \& Wiliams(2007)]{andrews07}Andrews, S.~M., \&
Williams, J.~P.\ 2007, \apj, 659, 705

\bibitem[Boss(1997)]{boss97}Boss, A.~P.\ 1997, Science, 276, 1836

\bibitem[Butler et al.(2006)]{butler06}Butler, R.~P., et al.\ 2006,
\apj, 646, 505 

\bibitem[Cameron(1978)]{cameron78}Cameron, A.~G.~W.\ 1978, Moon
and Planets, 18, 5

\bibitem[Chiang et al.(2008)]{chiang08}Chiang, E., Kite, E., Kalas,
P., Graham, J.~R., \& Clampin, M.\ 2009, \apj, 693, 734

\bibitem[Clarke et al.(2007)]{clarke07}Clarke, C.~J., Harper-Clark,
E., \& Lodato, G.\ 2007, \mnras, 381, 1543

\bibitem[Cotera et al.(2001)]{cotera01}Cotera, A.~S., et al.\ 2001,
\apj, 556, 958 

\bibitem[Durisen et al.(2007)]{durisen07}Durisen, R.~H., Boss, A.~P.,
Mayer, L., Nelson, A.~F., Quinn, T., \& Rice, W.~K.~M.\ 2007,
Protostars and Planets V, 607

\bibitem[Gammie(2001)]{gammie01}Gammie, C.~F.\ 2001, \apj, 553,
174

\bibitem[Greaves et al.(2008)]{greaves08}Greaves, J.~S., Richards,
A.~M.~S., Rice, W.~K.~M., \& Muxlow, T.~W.~B.\ 2008, \mnras,
391, L74

\bibitem[Haisch et al.(2001)]{haisch01}Haisch, K.~E., Jr., Lada,
E.~A., \& Lada, C.~J.\ 2001, \apjl, 553, L153 

\bibitem[Helling et al.(2000)]{helling00}Helling, C., Winters, J.~M.,
\& Sedlmayr, E.\ 2000, \aap, 358, 651

\bibitem[Ikoma et al.(2000)]{ikoma00}Ikoma, M., Nakazawa, K., \&
Emori, H.\ 2000, \apj, 537, 1013

\bibitem[Inaba et al.(2003)]{inaba03}Inaba, S., Wetherill, G.~W.,
\& Ikoma, M.\ 2003, Icarus, 166, 46 

\bibitem[Kalas et al.(2008)]{kalas08}Kalas, P., et al.\ 2008, Science,
322, 1345 

\bibitem[Kenyon \& Bromley(2008)]{kenyon08}Kenyon, S.~J., \& Bromley,
B.~C.\ 2008, \apjs, 179, 451

\bibitem[Kuiper(1951)]{kuiper51}Kuiper, G.~P.\ 1951, Proceedings
of the National Academy of Science, 37, 1

\bibitem[Marois et al.(2008)]{marois08}Marois, C., Macintosh, B.,
Barman, T., Zuckerman, B., Song, I., Patience, J., Lafreni{\`e}re,
D., \& Doyon, R.\ 2008, Science, 322, 1348 

\bibitem[Mayor \& Queloz(1995)]{mayor95}Mayor, M., \& Queloz, D.\ 1995,
\nat, 378, 355 

\bibitem[Pollack et al.(1996)]{pollack96}Pollack, J.~B., Hubickyj,
O., Bodenheimer, P., Lissauer, J.~J., Podolak, M., \& Greenzweig,
Y.\ 1996, Icarus, 124, 62 

\bibitem[Rafikov(2005)]{rafikov05}Rafikov, R.~R.\ 2005, \apjl,
621, L69

\bibitem[Rice et al.(2003)]{rice03}Rice, W.~K.~M., Armitage, P.~J.,
Bate, M.~R., \& Bonnell, I.~A.\ 2003, \mnras, 339, 1025

\bibitem[Stamatellos \& Whitworth(2009)]{stamatellos09}Stamatellos,
D., \& Whitworth, A.~P.\ 2009, \mnras, 392, 413 

\bibitem[Toomre(1964)]{toomre64}Toomre, A.\ 1964, \apj, 139, 1217

\bibitem[Vorobyov(2009)]{vorobyov09a}Vorobyov, E.~I.\ 2009, \apj,
692, 1609

\bibitem[Vorobyov \& Basu(2006)]{vorobyov06}Vorobyov, E.~I., \&
Basu, S.\ 2006, \apj, 650, 956 

\bibitem[Vorobyov \& Basu(2009)]{vorobyov09b}Vorobyov, E.~I., \&
Basu, S.\ 2009, \mnras, 393, 822
\end{thebibliography}
\end{document}